\documentclass[11pt]{article}

\usepackage{amsmath,amsfonts,amssymb,amsbsy,textcomp}
\usepackage{graphicx,color}
\usepackage{graphics}

\usepackage{amsmath, amssymb,graphicx}
\usepackage{epsfig}
\usepackage{verbatim}
\usepackage{amsfonts}
\usepackage{epsf}

\def\be{\begin{equation}}
\def\ee{\end{equation}}
\def\ba{\begin{eqnarray}}
\def\ea{\end{eqnarray}}
\def\bc{\begin{center}}
\def\ec{\end{center}}
\def\cO{{\mathcal{O}}}
\def\R{{\mathbb{R}}}

\def\nn{\nonumber}

\def\r2{{\sqrt{2}}}

\def\xp{x^{+}}
\def\xm{x^{-}}

\def\mpsu{{\mathfrak{psu}}}
\def\msu{{\mathfrak{su}}}
\def\cR{{\mathcal{R}}}
\def\w{\omega}

\begin{document}
\begin{titlepage}
\begin{flushright}
{\bf \today} \\
DAMTP-07-114\\
MAD-TH-07-12\\
arXiv:0712.1361 [hep-th] \\
\end{flushright}
\begin{centering}
%^{\rm full}
\vspace{.2in}

 {\Large {\bf Comments on the Boundary Scattering Phase}}

\vspace{.3in}

Heng-Yu Chen${}^{1,2,\dagger}$ and Diego H. Correa${}^{1,\ddagger}$ \\
\vspace{.2 in}
${}^{1}$DAMTP, Centre for Mathematical Sciences \\
University of Cambridge, Wilberforce Road \\
Cambridge CB3 0WA, UK \\
\vspace{.2in}
and \\
%\vspace{.2in}
\vspace{.1 in}
${}^{2}$Department of Physics, University of Wisconsin, \\
Madison, WI 53706, USA.\\
\vspace{.2in}
${}^{\dagger}$ {\tt hchen46@wisc.edu,}\quad ${}^{\ddagger}$ {\tt D.Correa@damtp.cam.ac.uk}
\vspace{.5in}

{\bf Abstract}

\vspace{.1in}

\end{centering}

We present a simple solution to the crossing equation
for an open string worldsheet reflection matrix, with boundaries preserving
a $SU(1|2)^2$ residual symmetry, which constrains the boundary dressing factor.
In addition, we also propose an analogous crossing equation for the dressing
factor where extra boundary degrees of freedom preserve
a $SU(2|2)^2$ residual symmetry.

%\vspace{.05in}
%\baselineskip=.3in
\end{titlepage}

\paragraph{}
The exciting discovery of integrable structures in the planar ${\cal
N}=4$ super Yang-Mills \cite{MZ,BKS,BS}\footnote{See also
\cite{BeisertThesis} for more comprehensive reference list.} has
allowed for the determination of the perturbative spectrum for
single trace operators. Particularly, in the limit of infinitely
long operators \cite{BMN,Staudacher,Beisert12,HM}, { some} exact
all-loop results, which are also valid in the strong coupling limit,
have been obtained. They allow precise comparisons with the energy
spectrum of the non-interacting closed strings in $AdS_5\times S^5$
carrying an infinitely large angular momentum, as prescribed by
AdS/CFT correspondence \cite{adscftmalda}. In such infinite limit,
physical content of the theory is the spectrum of asymptotic states
and their scattering matrix, and as a hallmark of integrable system,
it is highly constrained by the residual symmetry present. The
relevant asymptotic Bethe equations then arise from imposing
{periodic boundary  conditions}. In fact, the residual symmetry
proved to be sufficient to determine the two-particle scattering
matrix between the elementary excitations up to an overall scalar
factor \cite{Beisert12}. Such overall scalar factor, known as
``dressing factor'' in the literature, plays an important role in
interpolating the weak-strong coupling spectrum of the gauge/string
correspondence \cite{AFS, BHL, BES}. Unconstrained by the residual
symmetry, the dressing factor however needs to obey an extra
symmetry known as ``crossing'' imposed on the aforementioned
two-particle scattering matrix \cite{BHL,BES,Janik}. The crossing
symmetry  existing here also ensures the bound state of an
elementary excitation and its anti-particle\footnote{Sometimes
referred as the ``singlet'' states with respect to the residual
symmetry algebra \cite{Beisert12}. } scatters trivially
\cite{Beisert12}.
\paragraph{}
It is also possible to consider certain operators in ${\cal N}=4$
super Yang-Mills for which boundaries are present in the
corresponding spin chains. This class of operators typically arises
when one introduces extra D3-branes in the bulk where open strings
can end on. As such open string becomes infinitely extended, locally
near its end points, the system can be regarded as a semi-infinite
configuration and the reflection of bulk excitations on the boundary
has to be considered. Analogous to the bulk scattering matrix, the
boundary reflection matrix can be determined up to an overall scalar
factor by the residual symmetries preserving certain class of
integrable boundaries \cite{HMopen}\footnote{See \cite{Mann} for an
earlier discussion on integrable boundary conditions for open
strings in $AdS_5\times S^5$.}. Moreover the overall scalar,
sometimes referred as ``boundary dressing factor'', can again be
further restricted by analogous crossing symmetry conditions.

\paragraph{}
The main objective of this note is to study these crossing symmetry conditions in {two} different
cases depending on the residual symmetries preserved. {Here,} like in the usual relativistic integrable
theories \cite{GZ}, the boundary crossing symmetry condition relates the bulk and the boundary
scattering matrices \cite{HMopen}. By specifying the boundary crossing transformation in
relation to the bulk one, we shall point out a connection between bulk and boundary dressing
factors and propose a simple solution to the boundary crossing equation derived in \cite{HMopen}
for one of the two boundary conditions. Furthermore, we shall also present the crossing equation
constraining the boundary dressing factor for the other boundary condition.

\paragraph{}
In general, the elementary excitation in the asymptotic spin-chain
is known as ``magnon'', and it can be specified by its definite
momentum $p$, energy $E$ and  flavor. The magnons of different
flavor arise as they combine to form a short multiplet under the
residual symmetry group the spin chain ground state preserves. For
the infinite spin chain, as considered in \cite{Beisert12}, this is
given by a centrally extended supergroup $PSU(2|2)^2\ltimes R^3\cong
SU(2|2)^2\ltimes R^2$ which preserves the ground state consisting
exclusively of the complex adjoint scalar $Z$s. The energy of magnon
$E$ is then identified with one of the three central charges, and
can be related to the magnon momentum by the dispersion relation:
\begin{equation}
E^2-16g^2\sin^{2}\frac{p}{2}=1
\label{MagDisp}\,.
\end{equation}
Here the coupling $g$ is related to the 't Hooft coupling $\lambda$
as $16\pi^2g^2=\lambda$.
\paragraph{}
The unusual relation (\ref{MagDisp}) differs from the usual standard
relativistic one or the one for lattice vibrations ({\it e.g.}
phonons), but rather shares features of both. In fact, it describes
a complex torus with two non-trivial circles
\cite{BHL},\cite{Janik}. The first or ``real'' circle is given by
shifting the magnon momentum $p\to p+4\pi{\mathbb{Z}}$; whereas the
second or ``imaginary'' circle arises when we regard
$4g\sin\frac{p}{2}$ instead of $p$ as the relevant relativistic
momentum. Then for purely imaginary $4g\sin\frac{p}{2}$, the
equation $E^2+(i4g\sin\frac{p}{2})^2=1$ describes a unit circle.
Therefore the magnon energy $E$ and ``momentum'' $4g\sin\frac{p}{2}$
are in fact defined on a torus and can be uniformized and expressed
in terms of Jacobi  elliptic functions. We can also introduce an
alternative set of complex variables known as the ``spectral
parameters'' $x^{\pm}$, which are more convenient for our purposes
and we shall mostly use subsequently. They are related to the
momentum $p$ and the energy $E$ through:
\begin{equation}
e^{i p} = \frac{x^+}{x^-}\,,~~~~
E=\frac{1}{2}+\frac{ig}{\xp}-\frac{ig}{\xm}\,,
\label{DefSpecPara}
\end{equation}
and they are subject to the constraint:
\begin{equation}
\xp+\frac{1}{\xp}-\xm-\frac{1}{\xm}=\frac{i}{g}\,.
\label{xpxmconst}
\end{equation}
Combining (\ref{DefSpecPara}) and (\ref{xpxmconst}) directly
reproduces the dispersion relation (\ref{MagDisp}). Again the
$x^{\pm}$ should be uniformized on the torus and we shall later give
the explicit expressions in terms of Jacobi  elliptic functions.
\paragraph{}
The asymptotic spin chains with boundaries turn up when studying the
${\cal N}=4$ super Yang-Mills operators dual to D3-branes with open
string excitations. In the $AdS_5\times S^5$ background, such
D3-branes \cite{GG} can be chosen to wrap a holomorphic surface
within the $S^{5}$ \footnote{The $S^5$ is then given in our
conventions by the surface $|Z|^2+|W|^2+|Y|^2=1$.}. For instance,
for holomorphic surfaces $Y=0$ or $Z=0$ the D3-branes are
three-spheres of maximum size within $S^{5}$ and are usually called
``maximal giant gravitons''. The perturbative computations in the
dual field theory at weak coupling and the classical sigma model at
strong coupling, have both shown that maximal giant gravitons
provide integrable boundary conditions for the open string sigma
model \cite{BereVaz,AgaOY,Mann,HMopen}. Depending on the relative
orientations of the angular momentum of the giant graviton and the
open string ground state, two different cases can be considered. In
one case, we can take the $Y=0$ giant graviton and let the open
string ground state carry angular momentum along the $Z$ direction.
The operator corresponding to this configuration is \cite{BHLN}:
\begin{equation}
\cO_Y=\epsilon^{i_1,i_2,\dots, i_{N-1},i_N}_{j_1,j_2,\dots,
j_{N-1},j_N}Y^{j_1}_{i_1}Y^{j_2}_{i_2}
\dots Y^{j_{N-1}}_{i_{N-1}}(ZZ\dots ZZ)^{j_N}_{i_N}\,,
\label{DefOy}
\end{equation}
where $i_n,~j_n,~n=1,2\dots,N$ are the $SU(N)$ color indices.
Essentially, one replaces the last entry in the operator $\det Y$
which is dual to the $Y=0$ giant graviton by an infinite chain of
$Z$s. As explained in details in \cite{HMopen}, in order to preserve
both $Z$s and $Y$s in the ground state (\ref{DefOy}), the residual
symmetry group is reduced to $SU(1|2)^2\subset PSU(2|2)^2\ltimes
R^3$. The elementary magnon which transforms as ${\bf (2_B,2_F)}$
under each copy of $PSU(2|2)\ltimes R^3$ {}\footnote{The three
central charges are shared between both copies of $PSU(2|2)\ltimes
R^3$.} can now be expressed as an irreducible multiplet under each
copy of $SU(1|2)$.
\paragraph{}
Alternatively, we can take a $Z=0$ maximal giant graviton,
with an open string ground state carrying angular momentum
along the $Z$ direction \cite{HMopen}
\begin{equation}
\cO_Z(\chi_L,\chi_R)=\epsilon^{i_1,i_2,\dots, i_{N-1},i_N}_{j_1,j_2,\dots,
j_{N-1},j_N}Z^{j_1}_{i_1}Z^{j_2}_{i_2} \dots
Z^{j_{N-1}}_{i_{N-1}}(\chi_L ZZ\dots
ZZ\chi_R)^{j_N}_{i_N}\label{DefOz}\,.
\end{equation}
Boundary impurities are added to prevent the factorization into a
determinant plus a single trace \cite{BCV12}. In this case, the full
$PSU(2|2)^2\ltimes R^3$ residual symmetry group preserves the string
of $Z$s and the $Z$s in the determinant. The elementary magnons as
well as the boundary degrees of freedom will transform in the
fundamental representation (but carry different central charges) of
the extended $SU(2|2)^2$.
\paragraph{}
Let us first recall the situation without boundary conditions.
In such infinite asymptotic spin chain, the elementary magnons propagate
freely apart from pairwise scattering, the physical content of the
theory is therefore the two body scattering matrix $S(x_1,x_2)$.
With the presence of an additional boundary, one also needs to take
into account the scattering between the magnon and the boundary and
 encode such interaction in a ``reflection matrix $\cR(x)$''
{}\footnote{Sometimes in the literature one refers $S(x_1,x_2)$ as
the ``bulk'' scattering matrix to distinguish from the ``boundary''
scattering matrix ${\cal R}(x)$ \cite{GZ}.}. It was shown in \cite{Beisert12}  by
demanding the invariance of the scattering matrix $S(x_1,x_2)\equiv
S(x_1^{\pm},x_2^{\pm})$ under the residual symmetry algebra
$\mpsu(2|2)^2\ltimes {\mathbb{R}}^3$, that it can be constrained up to an
overall scalar. This is a hallmark of an integrable system and
$S(x_1,x_2)$ takes the following schematic form:
\begin{equation}
S_{\rm full}(x_1,x_2)=S^2_0(x_1,x_2)\left(\hat{S}_{\msu(2|2)}(x_1,x_2)\otimes
\hat{S}_{\msu'(2|2)}(x_1,x_2)\right)\,.
\label{BulkSmatrix}
\end{equation}
Here $\hat{S}_{\msu(2|2)}(x_1,x_2)$ and
$\hat{S}_{\msu'(2|2)}(x_1,x_2)$ are flavor dependent parts and are
uniquely fixed by each copy of $\msu(2|2)\ltimes {\mathbb{R}}^2$
respectively, and non-trivially satisfy the Yang-Baxter and
unitarity equations. This further confirms the integrable structure
of the theory. Whereas the remaining overall scalar factor
$S_0(x_1,x_2)^2$ is given by\footnote{Here we follow the conventions
as in \cite{BHL} and \cite{HMopen}.}
\begin{equation}
S_0(x_1,x_2)^2=
\frac{(x_1^+-x_2^-)(1-\frac{1}{x_1^-x_2^+})}{(x_1^--x_2^+)(1-\frac{1}{x_1^+x_2^-})}
\frac{1}{\sigma^2(x_1,x_2)}\,,
\label{DefS0}
\end{equation}
the function $\sigma(x_1,x_2)\equiv\sigma(x_1^{\pm},x_2^{\pm})$ is
usually referred to in the literature as the ``dressing factor''
\cite{Staudacher, AFS, BHL,BES}. To determine $\sigma(x_1,x_2)$
additional dynamical constraints, such as crossing symmetry in a
relativistic theory \cite{Zam} which interchanges
particle/anti-particle (and perhaps higher loop computations), are
required. Despite the non-standard dispersion relation for the
elementary magnon (\ref{MagDisp}), it was demonstrated in
\cite{Janik} that crossing symmetry can also be implemented provided
the dressing factor satisfies the crossing equations:
\begin{eqnarray}
\sigma(\bar x_1,x_2) {\sigma(x_1,x_2)} &\!\!=\!\!& \frac{x_2^-}{x_2^+}{f(x_1,x_2)}\,,
\label{cs1}
\\
\sigma(x_1,\bar  x_2) {\sigma(x_1,x_2)} &\!\!=\!\!&
\frac{x_1^+}{x_1^-}{f(x_1,x_2)}\,,
\label{cs2}
\end{eqnarray}
where $\bar{x}_{1,2}$ are the ``crossed'' spectral parameters which we
shall explain momentarily, and the function $f(x_1,x_2)$ is given by
\begin{eqnarray}
f(x_1,x_2) &\!\!\equiv\!\!&
\frac{(x^-_1-x^+_2)(1-1/x_1^+x_2^+)}{(x^-_1-x^-_2)(1-1/x_1^+x_2^-)}
=\frac{(x^-_1-x^+_2)(1-1/x_1^-x_2^-)}{(x^+_1-x^+_2)(1-1/x_1^+x_2^-)}\,.
\label{Defffunction}
\end{eqnarray}
These equations also need to be supplemented with the unitarity condition
\begin{equation}
\sigma(x_1,x_2) \sigma(x_2,x_1)=1\,.
\label{ucsig}
\end{equation}
Let us now explain how crossing symmetry can be implemented in the context
of magnon scattering following \cite{BHL, Janik}. As discussed earlier,
{the} magnon energy and momentum or equivalently the spectral parameters
$x^{\pm}$ are in fact defined on a torus. {Therefore} we can parametrize them, subject to the
dispersion relation (\ref{MagDisp}) or (\ref{xpxmconst}) in terms of
Jacobi elliptic functions \cite{BHL}:
\begin{equation}
p(z)=2{\rm am}(z,k)\,,~~
\sin\frac{p(z)}{2}={\rm sn}(z,k)\,,~~
E(z)= {\rm dn}(z,k)\,.
\label{DefellippE}
\end{equation}
Here the elliptic modulus $k=4ig$ {can} be taken fixed in a
conformal field theory, so that $p$ and $E$ can be regarded as
functions of the complex parameter $z$ called ``generalized
rapidity''\footnote{The terminology here is drawn from the
relativistic case, where the parametrization
$\epsilon=m\cosh\theta\,, p=m\sinh\theta\,,\theta\in{\mathbb{R}}$
satisfies the dispersion relation $\epsilon^2-p^2=m^2$.}. Simple
manipulation of elliptic {functions} identities shows that the
spectral parameters {are} given by,
\begin{equation}
x^{\pm}(z) = \frac{1}{4g}
\left(\frac{{\rm cn}(z,k)}{{\rm sn}(z,k)}\pm i\right)(1+{\rm
dn}(z,k))\,.
\label{Defellipxpxm}
\end{equation}
The two circles of the complex torus can be quantified by the shifts
$z\to z\pm2\omega_1$ and $z\to z\pm 2\omega_2$ with
\begin{equation}
\omega_1 = 2 {\rm K}(k^2)\,,\qquad \omega_2
= 2 i{\rm K}(1-k^2)-2 {\rm K}(k^2)\,,
\label{EllipPeriods}
\end{equation}
where ${\rm K}(k^2)$ is the complete elliptic integral of first
kind\footnote{Our convention is ${\rm
K}(k^2)=\int^{\pi/2}_{0}\frac{d\phi}{\sqrt{1-k^2\sin^2\phi}}$.}. For
elliptic modulus $k^2=-16g^2\in {\mathbb{R}}$, ${\rm
Im}(\omega_1)={\rm Re}(\omega_2)=0$ and $\omega_1$ and $\omega_2$
can be identified with the half-period of real and imaginary circles
respectively. The complex torus can then {be} defined on the generalized
rapidity $z$-plane by the domain $\left\{|{\rm Re}(z)|\le
\omega_1\,,|{\rm Im}(z)|\le \omega_2\right\}$. However, one should
stress that $(E(z),p(z))$ are not both real in the entire domain
and the function $S(z_1,z_2)$ should be regarded as an analytic continuation
of the S-matrix with real $E$ and $p$ (real values of the generalized rapidity) \cite{DHM,AF2}.

\paragraph{}
Now let us consider how $x^{\pm}(z)$ transform under the
translations on the $z$-plane. One can first verify that, along the
real axis, the shift $z\pm \omega_1$ leaves $x^{\pm}(z)$ invariant.
However along the imaginary axis the shift $z\pm \omega_2$
transforms $x^{\pm}(z)$ as
\begin{equation}
{\rm Crossing}~:~x^{\pm}(z)\to x^\pm
(z\pm\omega_2) =
\bar{x}^{\pm}(z)=\frac{1}{x^{\pm}(z)}\,.
\label{xcrossingTrans}
\end{equation}
Such transformation (\ref{xcrossingTrans}) is the ``crossing
transformation'' defined in terms of the spectral parameters given
in \cite{Janik}. Here the notation $\bar{x}^{\pm}(z)\equiv
\bar{x}^{\pm}$ are the crossed spectral parameters introduced
earlier in (\ref{cs1}) and (\ref{cs2}). In terms of the magnon
energy $E(z)$ and momentum $p(z)$, (\ref{xcrossingTrans})
corresponds to the transformation $(E(z),p(z))\to (-E(z),-p(z))$
which is precisely how crossing transformation acts in relativistic
theory, hence the terminology. One should note that in terms of the
spectral parameters $x^{\pm}$, the ``double crossing'', {\it i.e.}
applying  (\ref{xcrossingTrans}) twice, appears to be a trivial map.
However, on $z$-plane this is a non-trivial translation $z\to
z+2\omega_2$ going round the imaginary circle of the complex torus
once. A non-trivial monodromy can be seen from the fact that the
ratio
${\sigma(\bar{\bar{x}}_1,x_2)}/{\sigma(x_1,x_2)}={f(\bar{x}_1,x_2)}/{f(x_1,x_2)}\neq1$.
Also because of that, it turns out that the sign (orientation) in
the shift $z\to z\pm \omega_2$ implementing the crossing
transformation matters. The consistency of crossing transformations
(\ref{cs1})-(\ref{cs2}) with the unitarity condition demands us to
define the shifts in the two arguments of the bulk S-matrix with
``opposite'' signs\footnote{A more correct way to put it is that to
be consistent with the underlying Hopf-algebra, it is necessary to
act on the first argument with the anti-pode $\mathcal{S}$ and on
the second with the inverse $\mathcal{S}^{-1}$ \cite{Janik}, or vice
versa.}. That is, for the two arguments in
$\sigma(x_1,x_2)\equiv\sigma(x(z_1),x(z_2))$, there are two
equivalent ways in which crossing transformation can act:
\begin{eqnarray}
(+,-)&:& z_1\to z_1+\omega_2\,,
~~~~
z_2\to z_2-\omega_2\,,
\label{conve1}\\
(-,+)&:& z_1\to z_1-\omega_2\,,
~~~~
z_2\to z_2+\omega_2\,.
\label{conve2}
\end{eqnarray}
For definiteness, we will adhere to the convention (\ref{conve1})
for the rest of this note. A class of consistent crossing symmetric
dressing factors satisfying (\ref{cs1}), (\ref{cs2}) and (\ref{ucsig}) were
found in \cite{BHL}. Moreover a unique special solution which correctly
reproduces the weak coupling gauge theory results was further singled out in
\cite{BES}.

Finally, one notes the discrete parity transformation can be imposed on the
$z$-plane as $z\to-z$, such that:
\begin{equation}
{\rm Parity}~:~x^\pm (-z) = -{x^{\mp}(z)}\,.
\label{xParTrans}
\end{equation}
In terms of the magnon energy and momentum, this yields
$(E(-z),p(-z))= (E(z),-p(z))$ as expected.
\paragraph{}
The construction of the boundary reflection matrix ${\cal R}(x)$
proceeds in an almost identical way as for the bulk scattering
matrix $S(x_1,x_2)$ \cite{HMopen}. For the $Y=0$ case, its form can
again be constrained up to an overall scalar factor by demanding its
invariance under the residual $\msu(1|2)^2$ symmetry algebra.
Schematically, the result is:
\begin{equation}
\cR_{R}^{\rm full}(x)=\cR_{0R}^2(x)\hat{\cR}_{R}^{\msu(1|2)}(x)\otimes
\hat{\cR}_{R}^{\msu '(1|2)}(x)\,.
\label{DefRefMatrix}
\end{equation}
Here the subscript ${}_R$ denotes the scattering of  a magnon with the
right boundary\footnote{The reflection matrix $\cR_{L}(x)$ for left
boundary can be deduced from (\ref{DefRefMatrix}) using the parity
symmetry of the problem, {\it i.e}. $\cR_{L}(x(z))=\cR_{R}(x(-z))$ \cite{HMopen}.}.
$\hat{\cR}_{R}^{\msu(1|2)}(x)$ and $\hat{\cR}_{R}^{\msu'(1|2)}(x)$
denotes the flavor dependent parts, which are uniquely fixed to be
\begin{equation}
\hat{\cR}_{R}^{\msu(1|2)}(x)=\left(
\begin{array}{cccc}
- \frac{x^-}{x^+} & 0 & 0 & 0 \\
 0 & 1 & 0 & 0 \\
0 & 0 & 1 & 0 \\
0 & 0 & 0 & 1
\end{array}
\right)
\end{equation}
whereas $\cR_{0R}^2(x)$ is the boundary equivalent of the dressing factor.
\paragraph{}
To derive the relevant crossing symmetry equation for the boundary dressing factor $\cR_{0R}(x)$, one can
recall an alternative derivation for (\ref{cs1}) and (\ref{cs2}),
which demands the trivial scattering between the $\mpsu(2|2)\ltimes
{\R}^3$ singlet state, given explicitly by
\begin{equation}
\label{singlet0}
|{\mathbf 1}_{p,\bar p}\rangle =
\frac{\alpha}{\gamma_p\gamma_{\bar
p}\xi_p}\left(\frac{x^+}{x^-}-1\right) \epsilon_{ab}|{\cal Y}^-{\cal
Y}^-{\cal Z}^+\phi_p^a\phi_{\bar p}^b\rangle
+\epsilon_{\alpha\beta}|\psi_p^\alpha \psi_{\bar
p}^\beta\rangle\,,
\end{equation}
and an elementary magnon in the bulk \cite{Beisert12}.
\begin{figure}%[htb]
\begin{center}
\epsfxsize=3in\leavevmode\epsfbox{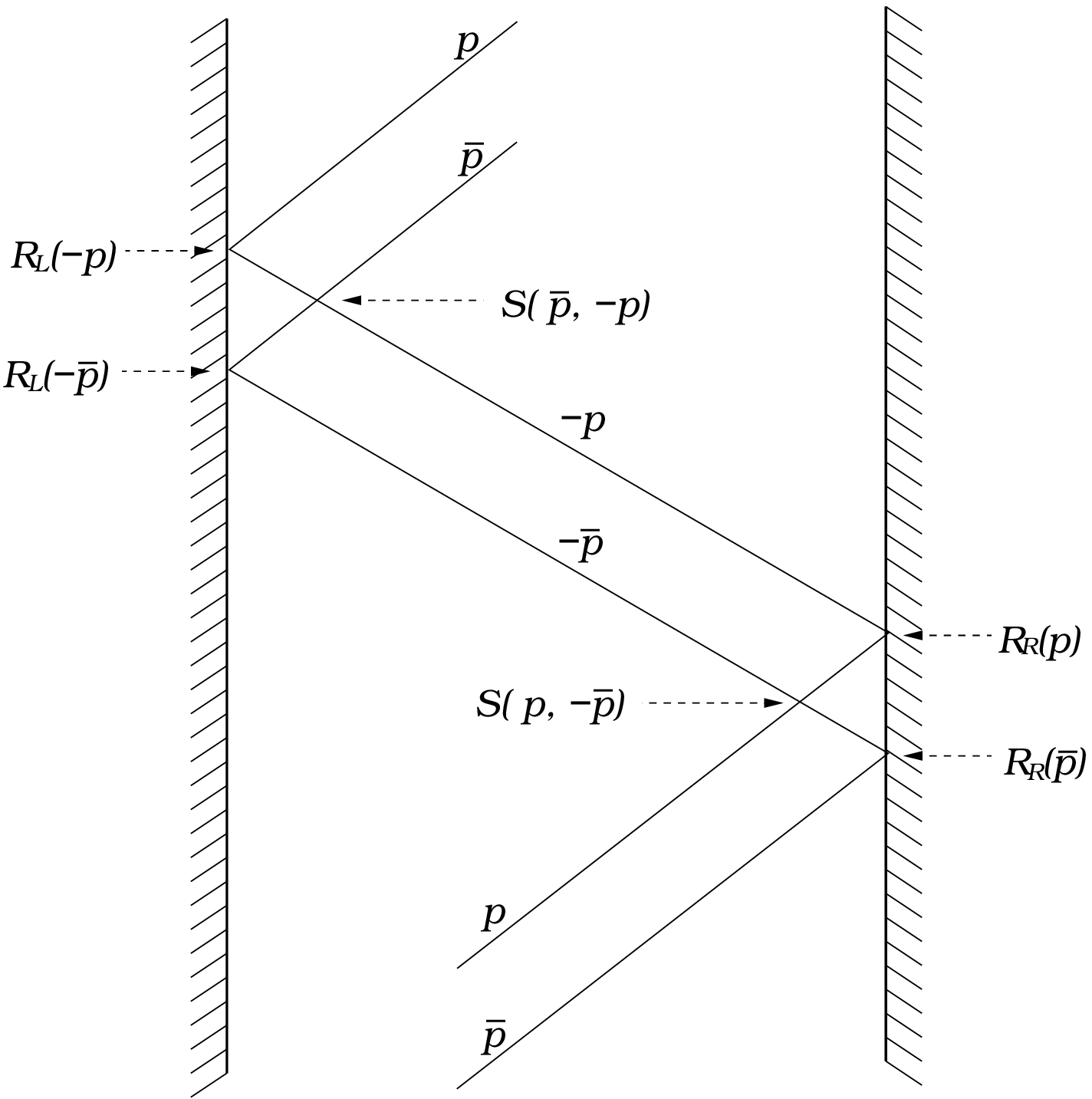}
\epsfxsize=3.35in\leavevmode\epsfbox{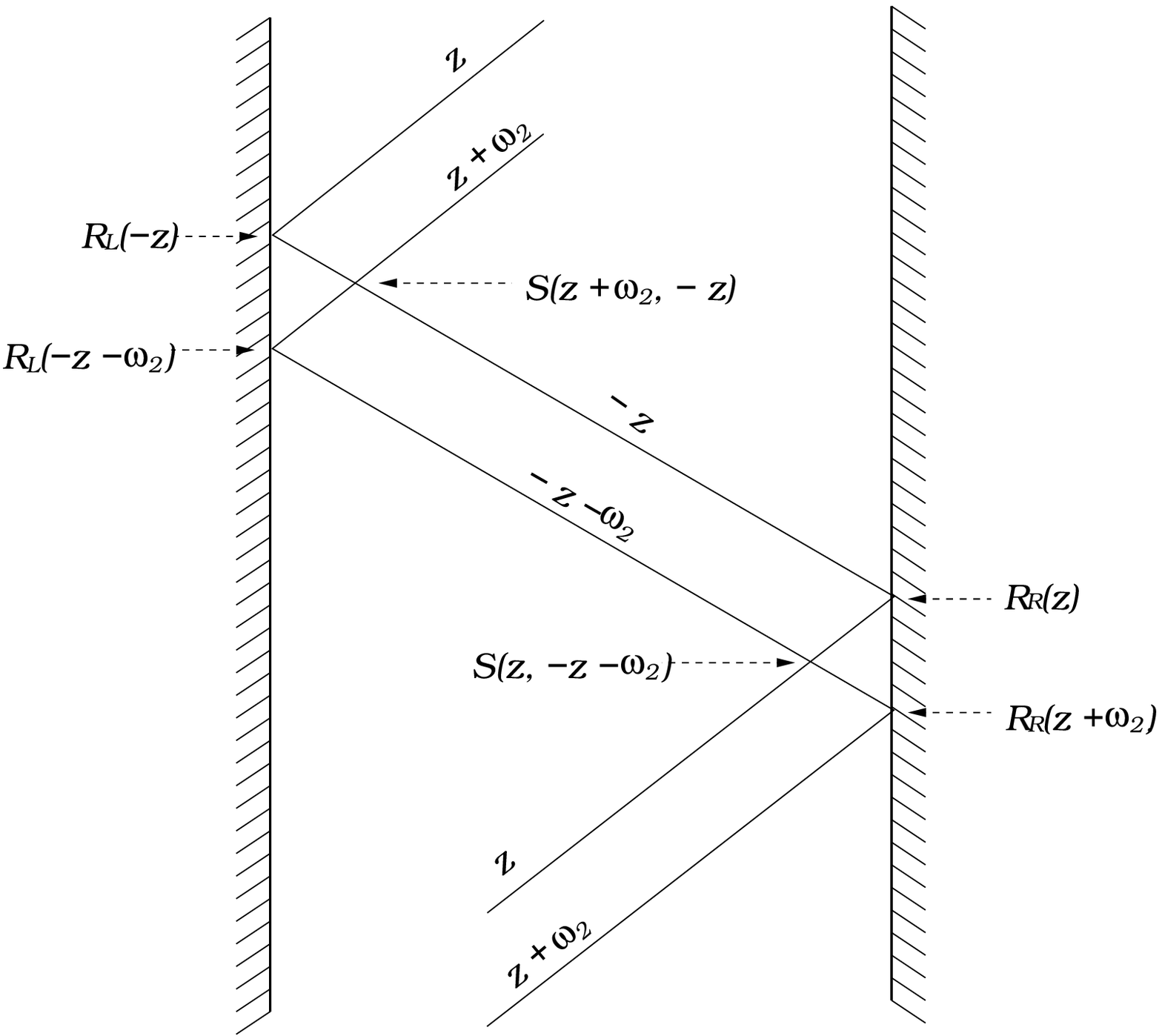}
\end{center}
\caption{Scattering of a singlet state, in terms of momenta (left)
and rapidities (right).} \label{fig}
\end{figure}
Similarly, if we demand that the singlet (\ref{singlet0}) scatters trivially
with the right boundary (as depicted in the lower right corner of each picture
in Figure \ref{fig}), the relevant scattering process can be described {as}
\begin{equation}
\label{rsca0}
|{\mathbf 1}_{p,\bar p}\rangle\to
{\cal R}_{R}(p)S(p,- \bar p){\cal R}_{R}(\bar p)|{\mathbf 1}_{p,\bar p}\rangle =
\frac{x^- +\frac{1}{x^-}}{x^+ +\frac{1}{x^+}}
S_0(p,- \bar p) {\cal R}_{0R}(x) {\cal R}_{0R}(\bar x)
|{\mathbf 1}_{- \bar p,-p}\rangle
\,.
\end{equation}
Symmetry under parity transformation of both, bulk and boundary
scattering matrix, implies that the reflection in the left boundary
will contribute with the same factor and revert the singlet to its
original orientation. This leads to a boundary crossing-symmetry
condition for the factor ${\cal R}_{0R}(p)$ \cite{HMopen}
\begin{equation}
{\cal R}_{0R}^2(p) {\cal R}_{0R}^2(\bar p)=
\frac{x^+ +\frac{1}{x^+}}{x^- +\frac{1}{x^-}}
\sigma^2(p,-\bar p)\,.
\label{ce}
\end{equation}
This equation again needs to obey the unitarity constraint:
\begin{equation}
{\cal R}_{0R}^2(p) {\cal R}_{0R}^2(- p)= 1.
\label{uc}
\end{equation}
As we shall show momentarily, it will be crucial to consider the generalized
rapidity $z$-plane coordinates and apply the crossing (\ref{xcrossingTrans})
transformation consistently in order to find the solution(s) to (\ref{ce}) and (\ref{uc}).
\paragraph{}
Essentially, one needs to specify how to implement the crossing
transformations in eq. (\ref{ce}), when writing it in terms of the generalized
rapidity $z$. The consistent choice of relative signs in the shifts would be
the one such that the action of the reflection is just flipping the sign
of the rapidity. This means that, as shown in Figure \ref{fig} for instance, the
argument of the boundary dressing factor for anti-particle $\bar p$,
has to be opposite in sign to the second argument of the bulk dressing
factor. Thus, if taking $\bar p$ as $p(z+\omega_2)$ in the LHS of (\ref{ce}),
the consistent choice is \footnote{Here we have used the
simplified notation $\sigma(z_1,z_2)\equiv \sigma(x(z_1),x(z_2))$,
${\cal R}_{0R}(z)\equiv {\cal R}_{0R}(x(z))$
and $x^{\pm}$ as $x^{\pm}(z)$ to avoid overlong expressions.}
\begin{equation}
{\cal R}_{0R}^2(z) {\cal R}_{0R}^2(z + \omega_2)=
\frac{x^+ +\frac{1}{x^+}}{x^- +\frac{1}{x^-}}
\sigma^2(z,-z-\omega_2)\,.
\label{cezcon}
\end{equation}

\paragraph{}
Before we present our proposed solution to the boundary crossing
equation (\ref{cezcon}), let us  first demonstrate that a different
choice of relative shift signs would have been inconsistent with
unitarity. Consider for example
\begin{equation}
{\cal R}_{0R}^2(z) {\cal R}_{0R}^2(z - \omega_2)=
\frac{x^+ +\frac{1}{x^+}}{x^- +\frac{1}{x^-}}
\sigma^2(z,-z-\omega_2)\,.
\label{cezincon}
\end{equation}
Without loss of generality we can write the boundary dressing factor
$\cR_{0R}^2(z)$ as
\begin{equation}
{\cal R}_{0R}^2(z) = F(z)\sigma(-z, z) \,,
\label{ansa}
\end{equation}
where $\sigma(z_1,z_2)$ satisfies (\ref{cs1}), (\ref{cs2}) and (\ref{ucsig}).
There are a few ways to motivate this. First in the so-called ``Giant Magnon regime'' \cite{HM,DHM},
the strong coupling computation
using classical string theory \cite{HMopen} shows that
\begin{equation}
{\cal R}_{0R}^2(z) = e^{ig\theta_{0}(-z,z)+{\cal O}(1)}\,,
\label{hmsc}
\end{equation}
where $\log \sigma(z_1,z_2)=ig\theta_0(z_1,z_2)+{\mathcal{O}}(1)$ is
the leading dressing phase calculated originally in \cite{AFS,HM}.
The ans\"atz (\ref{ansa}) correctly captures the strong coupling
result (\ref{hmsc}) provided $\log F(z)\to {\mathcal{O}}(1)$ as
$g\to\infty$. Second, the factor $\sigma(p,-\bar{p})$ on the RHS of
(\ref{ce}) can be easily re-written in terms of $\sigma(-z,z)$
provided consistent crossing convention is taken, {then} we are the
left with conditions for $F(z)$.
\paragraph{}
With the ans\"atz (\ref{ansa}), the boundary crossing equation (\ref{cezincon})
would be written as
\begin{equation}
F(z)F(z-\w_2)\sigma(-z,z)\sigma(-z+\omega_2, z-\omega_2)
= \frac{x^+ +\frac{1}{x^+}}{x^- +\frac{1}{x^-}}
\sigma^2(z,-z-\omega_2)\,.
\label{ceb}
\end{equation}
The successive application of (\ref{cs1})-(\ref{cs2}), allows one to write
\begin{equation}
\frac{\sigma(-z, z) \sigma(-z+\omega_2,
z-\omega_2)}{\sigma^2(z,-z-\omega_2)}=
\frac{f(-z+\omega_2,z)}{f(z,-z)^2 f(-z,z)} = \frac{(x^+
+\frac{1}{x^+})^3(x^- +\frac{1}{x^-})(x^+ + x^-)^4}{16(1+x^+ x^-)^4}
\,.
\label{wrongcrossing}
\end{equation}
To satisfy (\ref{ceb}) we would need to impose
\begin{equation}
F(z)F(z-\omega_2)= \frac{16(1+x^+ x^-)^4}{(x^+
+\frac{1}{x^+})^2(x^- +\frac{1}{x^-})^2(x^+ + x^-)^4}\,.
\label{cebf}
\end{equation}
It is important to notice that the RHS of (\ref{cebf}) is
crossing invariant. As a consequence,
\begin{equation}F(z-2\omega_2)=
F(z)\,,
\label{wrongcrossing2}
\end{equation}
{\it i.e.} the non-trivial factor due to double crossing the
argument of ${\cal R}_{0R}(z)$ is completely accounted by the shift
on the factor $\sigma(-z, z)$. Now if we apply the parity
transformation $z\to -z$ to (\ref{cebf}) and use unitarity condition
$F(z)F(-z)=1$, we get
\begin{equation}
\frac{F(-z-\omega_2)}{F(z)}= \frac{16(1+x^+ x^-)^4}{(x^+
+\frac{1}{x^+})^2(x^- +\frac{1}{x^-})^2(x^+ + x^-)^4}\,,
\label{cebfc1}
\end{equation}
whereas imposing unitarity only on the original (\ref{cebf}) gives
\begin{equation}
\frac{F(z)}{F(-z+\omega_2)}= \frac{16(1+x^+
x^-)^4}{(x^+ +\frac{1}{x^+})^2(x^- +\frac{1}{x^-})^2(x^+ +
x^-)^4}\,.
\label{cebfc2}
\end{equation}
However (\ref{wrongcrossing2}) would then tell us
\begin{equation}
\frac{16(1+x^+ x^-)^4}{(x^+ +\frac{1}{x^+})^2(x^-
+\frac{1}{x^-})^2(x^+ + x^-)^4}=1\,,
\end{equation}
which is clearly not the case for arbitrary $x^\pm$ ! This lead us conclude that
equations (\ref{cebfc1}) and
(\ref{cebfc2}) are in obvious contradiction.
This means that with the choice of shift signs (\ref{cezincon}), boundary crossing
symmetry and unitarity would be inconsistent with each other.

\paragraph{}
Now let us see that when adopting the other choice of shift signs
(\ref{cezcon}), which we argued to be the consistent one, both boundary
crossing equation and unitarity condition can be simultaneously solved
with ease. Using once again the ans\"atz (\ref{ansa}) we obtain
\begin{equation}
F(z)F(z+\w_2)\sigma(-z, z)\sigma(-z-\omega_2, z+\omega_2)
= \frac{x^+ +\frac{1}{x^+}}{x^- +\frac{1}{x^-}} \sigma^2(z,-z-\omega_2)\,.
\label{cea}
\end{equation}
According to (\ref{cs1}) and (\ref{cs2}),
\begin{equation}
\frac{\sigma(-z, z) \sigma(-z-\omega_2, z+\omega_2)}{\sigma(z,-z-\omega_2)^2}=
\frac{1}{f(z,-z-\omega_2)f(z+\w_2,-z-\w_2)}
= \frac{x^+ +\frac{1}{x^+}}{x^- +\frac{1}{x^-}}\,.
\end{equation}
Thus we obtain the following crossing symmetry and unitarity conditions imposed on $F(z)$
\begin{eqnarray}
&& F(z)F(z+\omega_2)=1\,,\nn\\
&& F(z)F(-z)=1\,,
\end{eqnarray}
The system can be solved generally by $F(z)=\pm\exp(if_{odd}(p))$ where $f_{odd}(p)$ is an
arbitrary odd function of the magnon momentum $p$, moreover at strong coupling when $g\to\infty$,
$\log F(z)\to {\mathcal{O}}(1)$.  By further comparing with the weak coupling expression in
(4.60) of \cite{HMopen} for $\cR_{0R}^{2}(z)$, this further requires that $F(z)\to-\exp(2ip)$
when $g\to 0$. Thus, the simplest solution for $\cR_{0R}^2(z)$ satisfying
the $Y=0$ case crossing equation (\ref{ce}) and in agreement with the known strong/weak
coupling results is:
\begin{equation}
{\cal R}_{0R}^2(z) = -\exp(2ip)\sigma(-z, z)\equiv
-\exp(2ip(z))\sigma(x(-z),x(z))\,.
\label{crosssolution}
\end{equation}
Let us briefly comment on the uniqueness of (\ref{crosssolution}).
Recall that the factor $\sigma(p,-\bar{p})$ entering in the RHS of
(\ref{ce}) is unambiguously interpreted as the unique bulk dressing
factor identified in \cite{BES}. We then showed that $\cR_{0R}^2(z)$
is given in terms of $\sigma(-z,z)$ satisfying
(\ref{cs1})-(\ref{cs2}). Therefore, for consistency, we should also
interpret $\sigma(-z,z)$ to have exactly the same functional form as
the unique bulk dressing factor identified in \cite{BES}. A good way
to verify our solution would be to explicitly calculate the leading
semi-classical $\frac{1}{g}$ correction in $\cR_{0R}^2(z)$ by
generalizing the worldsheet approach in \cite{CDM} to the situation
with open boundary conditions.

\paragraph{}
We end this note by considering also the crossing equation for the
case $Z=0$. Here the form of the boundary reflection matrix
${\cR}(x)$ is again fixed up to an overall scalar factor
$\cR_{0R}(x)$ by demanding its invariance under the residual
$\msu(2|2)^2$ symmetry algebra \cite{HMopen},
\begin{equation}
\cR_{R}^{\rm full}(x)=\cR_{0R}^2(x)\hat{\cR}_{R}^{\msu(2|2)}(x)\otimes \hat{\cR}_{R}^{\msu '(2|2)}(x)\,.
\label{DefRefMatrix2}
\end{equation}
Because of the extra boundary degree of freedom, $\hat{\cR}_{R}^{\msu(2|2)}(x)$ and
$\hat{\cR}_{R}^{\msu'(2|2)}(x)$ are now $16\times 16$ matrices. They are non-diagonal
and their complexity is similar to that of $\hat{S}_{\msu(2|2)}(x_1,x_2)$. The
explicit expressions for their components are presented in (3.42)-(3.46) of \cite{HMopen}
and they have also been checked to satisfy the boundary Yang-Baxter equations.

\paragraph{}
As done in \cite{HMopen} for the case $Y=0$, we can use the
reflection of the singlet state (\ref{singlet0}) to derive a
relevant crossing symmetry condition. Now from the factorizability,
the scattering processes displayed in the right lower corner of
Figure 1 consist of the action  of three $16\times 16$ dimensional
scattering matrices. After a computationally intense calculation we
obtain, regardless of the flavor of the boundary impurity, the
following result for reflection of the singlet state:
\begin{equation}
\label{rsca02}
{\cal R}_{R}(p)S(p,- \bar p){\cal R}_{R}(\bar p)|{\mathbf 1}_{p,\bar p}\ \chi_R\rangle =
\frac{x^- +\frac{1}{x^-}}{x^+ +\frac{1}{x^+}} h_B(p) S_0(p,- \bar p) {\cal R}_{0R}(x) {\cal R}_{0R}(\bar x)
|{\mathbf 1}_{- \bar p,-p}\ \chi_R\rangle\,,
\end{equation}
where
\begin{equation}
h_B(p) = \frac{x^+(x_B-x^-)}{x^-(x_B-x^+)} \frac{1+(x_B
x^-x^+)^2}{(1-(x_B x^+)^2)(1-x^-x^+)}\, ,
\end{equation}
and $x_B = \frac i{2g}(1+\sqrt{1+4g^2} )$. This leads to a
crossing-symmetry condition for the boundary dressing factor ${\cal
R}_{0R}(p)$
\begin{equation}
{\cal R}_{0R}^2(p) {\cal R}_{0R}^2(\bar p)=
\frac{x^+ +\frac{1}{x^+}}{x^- +\frac{1}{x^-}}
 \frac{\sigma^2(p,-\bar p)}{h_B(p)^2}\,.
\label{ce2}
\end{equation}
As a simple check for (\ref{ce2}), one notices that in the leading $g\to\infty$ expansion
(or more precisely, the giant magnon limit $x^{\pm}\sim e^{\pm ip/2}$.),
%the crossing and parity transformations/equations coincide, this demands that
$h_B(x)\to\cO(1)$ and then the classical dressing factor
%in this limit, happily it does. Or equivalently one can also check that the classical dressing factor
%
\begin{equation}
{\cal R}_{0R}^2(z) =e^{ig\theta_{0}(-z,z)+2ig\theta_{0}(\omega_1/2,z)+{\cal O}(1)}\,,
\label{hmsc2}
\end{equation}
deduced in \cite{HMopen} using the ``method of images'' satisfies the classical limit of the crossing
equation (\ref{ce2}). Up to the factor $1/h_B(p)^2$, one recognizes the crossing symmetry equation of the
$Y=0$ case, which {we showed} can  be simply solved by $\sigma(-z,z)$.
Under crossing and unitarity conditions, the naive ansatz ${\cal R}_{0R}^2(z) = F(z)\sigma(-z, z)$ yields
\begin{eqnarray}
&& F(z)F(z+\omega_2)=\frac{1}{h_B(p(z))^2}\,,\nn\\
&& F(z)F(-z)=1\,.
\end{eqnarray}
and we have not attempted to solve this set of equations. One should also notice that
$h_B(p(z))^2$ is not crossing invariant, therefore the function $F(z)$ still has a
non-trivial monodromy, {\it i.e.} $F(z+2\omega_2)\neq F(z)$, and then
a non trivial crossing condition remains to be solved.

\paragraph{}
As a first step however, it would be very desirable to obtain the
leading semiclassical correction to the boundary dressing factor in
this $Z=0$ case as in its closed string counterpart \cite{HL} and
test if crossing symmetry condition (\ref{ce2}) is obeyed (c.f.
\cite{AF}). In particular, in contrast with the case $Y=0$, here the
boundary analogue of the magnon bound states exists \cite{HMopen,
Dorey}. Therefore, it would  be interesting to generalize the
constructions of multi-soliton solutions in \cite{CDM,CDO1}, also
their classical \cite{CDO2} and semiclassical scatterings \cite{CDM}
to the case with boundaries. Here the fluctuations around the
soliton background would need to be subject to appropriate boundary
conditions, this direction is currently under investigation.

\paragraph{}
%\subsection*{Acknowledgments}
The authors would like to thank N. Dorey for numerous
discussions and comments on the draft. They would also like to thank R. Janik,
N. Mackay, N. Mann, R. Matos, K. Okamura for useful discussions and D.~Hofman for reading and commenting the draft. HYC
is supported in part by DOE grant DE-FG-02-95ER40896 and the funds
from University of Wisconsin-Madison. DHC is supported by  PPARC
grant ref. PP/D507366/1.

%%%%%%%%%%%%%%%%%%%%%%%%%
%%%%%%%%%%%%%%%%%%%%%%%%%

\end{document}